\documentstyle[aps,preprint,epsfig,amsmath]{revtex}

\def \beq {\begin{equation}}
\def \eeq {\end{equation}}

\begin{document}

\draft

\title{Conformal quantum
effects and the anisotropic singularities of scalar-tensor theories of gravity}

\author{
E. Gunzig$^{1}$,
Alberto Saa$^{2}$ \footnote{e-mails: 
{ 
egunzig@ulb.ac.be, 
asaa@ime.unicamp.br}}}

\address{1)
RggR, Universit\'e Libre de Bruxelles, 
CP 231, 1050 Bruxelles, Belgium.
}

\address{2)
IMECC -- UNICAMP,
C.P. 6065, 13081-970 Campinas, SP, Brazil.}

\maketitle

\begin{abstract}
We show that the inclusion of a term $C_{abcd}C^{abcd}$
in the action can remove the 
recently described anisotropic singularity
occurring on the hypersurface $F(\phi)=0$ of
scalar-tensor theories of gravity  of the type
$$
S=\int d^4x \sqrt{-g}\left\{F(\phi)R - \partial_a\phi\partial^a\phi
-2V(\phi) \right\},
$$
preserving, by construction, all of their isotropic solutions.
We show that, in principle, a 
higher order term of this type can arise
from considerations about the renormalizability of the
semiclassical approach to the theory. Such result brings 
again into consideration the quintessential models recently
proposed based in a conformally coupled scalar field
($F(\phi)=1-\frac{1}{6}\phi^2$) with potential $V(\phi)=\frac{m}{2}\phi^2
-\frac{\Omega}{4}\phi^4$, that have been discharged as unrealistic 
precisely by their anisotropic instabilities on the hypersurface
$F(\phi)=0$.

\end{abstract}

\section{Introduction}
In \cite{PRD}, it was proposed a
quintessential model corresponding to the homogeneous and
isotropic solutions of the cosmological model described by the action:
\beq
\label{act}
S=\int d^4x \sqrt{-g}\left\{F(\phi)R - \partial_a\phi\partial^a\phi
-2V(\phi) \right\},
\eeq
with $F(\phi)=1-\frac{1}{6}\phi^2$,  
the so-called conformal
coupling, and $V(\phi)=\frac{m}{2}\phi^2-\frac{\Omega}{4}\phi^4$.
Some novel and interesting dynamical behaviors were identified: 
superinflation regimes, 
a possible 
avoidance of big-bang singularities through classical birth of the universe 
from empty Minkowski space, spontaneous entry into and exit from inflation,
and a cosmological history suitable 
for describing quintessence. 
The next natural 
step was\cite{PRD2} the analysis of the robustness of these results
against small
perturbations in initial conditions and in the model itself. 
By considering general coupling $F(\phi)$ models as in \cite{G},
we generalize previous results\cite{Starobinski,Futamase,s2} and
identify two kinds of dynamically unavoidable singularities in
the models described by (\ref{act}).

The first one appears only in the anisotropic case and
corresponds to the hypersurfaces $F(\phi)=0$. It is a direct
generalization of the Starobinski singularities of conformally
coupled anisotropic solution\cite{Starobinski}. It
implies that the homogeneous
and isotropic solutions 
passing from the $F(\phi)>0$ to the  $F(\phi)<0$ region in the
model described in \cite{PRD} are
extremely unstable against anisotropic perturbations, challenging
its proposal as a quintessential model. 
The second type of singularity corresponds to $F_1(\phi)=0$, with
\beq
\label{f1}
F_1(\phi) = F(\phi)+\frac{3}{2}\left(F'(\phi)\right)^2,
\eeq
and it is present even for the homogeneous and isotropic cases. 
Futamase and co-workers\cite{Futamase} identified both
singularities
in the context of chaotic inflation in $F(\phi)=1-\xi\phi^2$ theories
(See also \cite{s2}).
The first singularity is always present for $\xi>0$ and the second one for 
$0<\xi<1/6$.
The conclusions of \cite{PRD2} are, however, more general since we treat the case of
general $F(\phi)$ and  $V(\phi)$ 
and our results are based on the analysis of
true geometrical invariants. The main result  is that the
system governed by (\ref{act}) is {\em generically} singular on both
hypersurfaces $F(\phi)=0$ and $F_1(\phi)=0$. 

The anisotropic singularity occurring on
$F(\phi)=0$ was the major obstacle in the developing of the model
proposed in \cite{PRD}. Note that   singularities
of the second type
are absent in the conformally coupled case. 
Although the isotropic solutions are always regular on the
hypersurface $F(\phi)=0$, any small deviation of isotropy will have
catastrophic consequences, leading to a spacetime singularity
 in a finite time. Even a
very small amount of anisotropy will be hugely amplified, feeding
the energy content of the scalar field $\phi$ and increasing the
spacetime curvature toward a true singularity.
Our purpose here
is to show that the inclusion of the higher order term
$C_{abcd}C^{abcd}$ can eliminate this anisotropic singularity
preserving, by construction, all isotropic solutions. Moreover,
we will see that a quantum counterterm precisely of this form
can arise from considerations about the renormalizability of
the semiclassical theory described by (\ref{act}).

The next section presents a brief review of the
geometric nature of the anisotropic singularity.
Section \ref{sec3} discusses the possible appearance of the
conformal counterterm $C_{abcd}C^{abcd}$. Its dynamical
implications are presented in Section \ref{sec4}. The last
section presents some concluding remarks.

\section{The singularity}

The equations derived from the 
action (\ref{act})
are the Klein-Gordon equation
\beq
\label{kg}
\Box\phi - V'(\phi) +\frac{1}{2}F'(\phi)R=0,
\eeq
and the Einstein equations
\begin{eqnarray}
\label{ee}
F(\phi)G_{ab} &=& (1+F''(\phi))\partial_a\phi\partial_b\phi \nonumber \\ &-& 
\frac{1}{2}g_{ab}\left[ (1+2F''(\phi))\partial_c\phi\partial^c\phi 
+2V(\phi)\right] - F'(\phi)\left(g_{ab}\Box\phi - \nabla_a\phi\nabla_b\phi 
\right).
\end{eqnarray}
We considered the simplest anisotropic homogeneous cosmological
model, the Bianchi type I, whose spatially flat metric is given by
\beq
\label{metric}
ds^2 = -dt^2 + a^2_1(t)dx^2 + a^2_2(t)dy^2 + a^2_3(t)dz^2. 
\eeq 
The dynamically relevant quantities here are 
$ H_i = \dot{a}_i/a$, $i=1,2,3$.
For such a metric and a homogeneous scalar field $\phi=\phi(t)$ 
Eq. (\ref{ee}) can be written as
\begin{eqnarray}
\label{ec}
F(\phi)G_{00} &=& \frac{1}{2}\dot{\phi}^2 + V(\phi) - F'(\phi)\left( 
H_1+H_2+H_3\right)\dot{\phi}, \\
\label{e1}
\frac{1}{a^2_1}F(\phi)G_{11} &=& \frac{1+2F''(\phi)}{2}\dot{\phi}^2 -
V(\phi) - F'(\phi)\left( H_1\dot{\phi} + V'(\phi) -\frac{F'(\phi)}{2}R\right),
\\\label{e2}
\frac{1}{a^2_2}F(\phi)G_{22} &=& \frac{1+2F''(\phi)}{2}\dot{\phi}^2 -
V(\phi) - F'(\phi)\left( H_2\dot{\phi} + V'(\phi) - \frac{F'(\phi)}{2}R\right),
\\\label{e3}
\frac{1}{a^2_3}F(\phi)G_{33} &=& \frac{1+2F''(\phi)}{2}\dot{\phi}^2 -
V(\phi) - F'(\phi)\left( H_3\dot{\phi} + V'(\phi) - \frac{F'(\phi)}{2}R\right).
\end{eqnarray}
It is quite simple to show that Eqs. (\ref{e1})-(\ref{e3}) are
not compatible, in general,
 on the hypersurface $F(\phi)=0$.  Subtracting
(\ref{e2}) and (\ref{e3}) from (\ref{e1}) we have, on 
such hypersurface, respectively,
\beq\label{sub}
F'(\phi)(H_1-H_2)\dot{\phi} = 0,\ {\rm and\quad }
F'(\phi)(H_1-H_3)\dot{\phi} = 0.
\eeq
Hence, 
they cannot be fulfilled in general for anisotropic metrics. This is
the origin of the anisotropic singularity.
Using the
new dynamical variables $p=H_1+H_2+H_3$, $q=H_1-H_2$, and 
$r=H_1-H_3$, Einstein Eqs. can be cast in the form
\begin{eqnarray}
\label{ec1}
E(\phi,\dot{\phi},p,q,r)&=&-\frac{1}{3}F(\phi)\left( 
p^2 + qr - q^2 - r^2
\right) + \frac{\dot{\phi}^2}{2} + V(\phi) - pF'(\phi)\dot{\phi} = 0,\\
\label{eq}
\dot{q} &=&  - \left(p+\frac{F'(\phi)}{F(\phi)}\dot{\phi}\right)q, \\
\label{er}
\dot{r} &=&  - \left(p+\frac{F'(\phi)}{F(\phi)}\dot{\phi}\right)r, \\
\label{ep}
-2F_1(\phi)\dot{p} &=& (F(\phi)+2F'(\phi)^2)p^2+
\frac{3}{2}(1+2F''(\phi))\dot{\phi}^2 - 3V(\phi) - 3 F'(\phi)V'(\phi) 
\nonumber \\
 & & - p\dot{\phi}F'(\phi) 
  + (F(\phi)+F'(\phi)^2)(q^2 + p^2 - qr)
\end{eqnarray}
A closer analysis of Eqs. (\ref{eq})-(\ref{er}) reveals the 
presence of the
singularity. In general, the right-hand side of these equations
diverge for $F(\phi)=0$. One can check that
this divergence is indeed related to real geometrical singularity
by considering the Kretschman scalar $I=R_{abcd}R^{abcd}$\cite{PRD2}.
Furthermore, it is dynamically unavoidable since the hypersurface
$F(\phi)=0$ has always an attractive side.

\section{Renormalization and quantum counterterms}
\label{sec3}

The idea of incorporating vacuum semiclassical 
effects into gravity has a long
history, and a good set of references is presented in
\cite{Birrel,Buchbinder}.
Zeldovich was the first to propose\cite{Zeldovich}, in 1967, 
that a cosmological
constant term could arise from quantum considerations of matter.
Yet in the sixties, in a set of seminal works, Parker 
considered\cite{Parker} 
the effect of the creation
of particles in a expanding universe, and discussed the possible
backreaction, opening the discussion of anisotropy damping and
avoidance of the initial singularity due to quantum 
corrections\cite{HartleHu}.

A semiclassical treatment of the model described by (\ref{act})
with $F(\phi)=1-\xi\phi^2$ and $V(\phi)=\frac{m}{2}\phi^2-
\frac{\Omega}{4}\phi^4$, 
where, hereafter,
by semiclassical one means that $\phi$ is quantized on a
classical gravitational background, requires the inclusion of
higher orders counterterms to ensure the renormalization of
the theory. These terms are\cite{Birrel,Buchbinder}
\beq
\label{vac}
S_{\rm vac} = \int d^4x\sqrt{-g}\left( 
\alpha_1 R^2 + \alpha_2 R_{ab}R^{ab}
+\alpha_3 R_{abcd}R^{abcd}
+\alpha_4\Box R\right).
\eeq
The quantum 
divergences of the semiclassical
theory can be removed by the renormalization
of the constants $\alpha_{1,2,3,4}$ and the Newtonian constant $G$.
In fact, the full set of quantities affected by the renormalization
includes
the matter field, its mass $m$ and selfcoupling constant $\Omega$,
the non-minimal coupling constant $\xi$ and  yet a cosmological 
constant. All these quantities are, in principle, subject to some
quantum running, and indeed some  
models to describe the reacceleration of the universe
have been recently proposed based on vacuum quantum 
effects\cite{VCDM,Shapiro}

The last counterterm in (\ref{vac}) does not contribute to the
classical dynamics, since it is merely a total divergence. In four
dimensions, we have
\beq
\label{conf}
C_{abcd}C^{abcd}=R_{abcd}R^{abcd}-2R_{ab}R^{ab}+
\frac{1}{3}R^2.
\eeq
Hence, it is possible, in principle, to combine
$\alpha_1$, $\alpha_2$ and $\alpha_3$ in order to have the
desired counterterm. We leave the issue of the naturalness of this
finely tunned choice to the last section.
The Weyl tensor $C_{abcd}$ vanishes identically for isotropic
spacetimes, and hence any contribution from this counterterm
would affect only the anisotropic case by construction, 
preserving all isotropic
solutions. The task of calculating the 
variations of the conformal counterterm $C_{abcd}C^{abcd}$ 
with respect to
the metric is simplified if
one recalls that in four dimensions the Gauss-Bonnet term
\beq
\label{euler}
E = R_{abcd}R^{abcd} + R^2 - 4R_{ab}R^{ab}
\eeq
has identically 
vanishing Euler-Lagrange equations and, hence, does not contribute
to the classical dynamics too, implying that 
the conformal counterterm is
dynamically equivalent to the term
$R_{abcd}R^{abcd}/2-R^2/6$. Thus,
with the inclusion of the conformal counterterm in the model
proposed in \cite{PRD}, the resulting dynamics are governed
by the action
\beq
\label{act1}
S=\int d^4x \sqrt{-g}\left\{\left(
1-\frac{1}{6}\phi^2 
\right)R + \alpha \left( \frac{1}{2}R_{abcd}R^{abcd} - \frac{1}{6}R^2
\right) - \partial_a\phi\partial^a\phi
-2V(\phi) \right\},
\eeq
where $V(\phi)=\frac{m}{2}\phi^2-\frac{\Omega}{4}\phi^4$ and
$\alpha$ is a parameter typically small when compared to $1/G$.

\section{The dynamics}
\label{sec4}

We study here the dynamics governed by the action (\ref{act1}).
The Klein-Gordon equation (\ref{kg}) is not affected by the
new term. It is clear,  however, that new higher order terms
will appear in the left handed side of Einstein equations
(\ref{ec})-(\ref{e3}). The new terms come from the tensors
\beq
Q_{ab} = \frac{1}{\sqrt{-g}}\frac{\delta}{\delta g^{ab}} 
\int d^4x\sqrt{-g} R_{abcd}R^{abcd} 
\eeq
and
\beq
S_{ab} = \frac{1}{\sqrt{-g}}\frac{\delta}{\delta g^{ab}} 
\int d^4x\sqrt{-g} R^2.
\eeq
We calculate the tensor $Q_{ab}$, $S_{ab}$ and $G_{ab}$  by
recalling that for the metric (\ref{metric}) one has
\beq
R = 2\left(\dot{H}_1 + \dot{H}_2 + \dot{H}_3
+H^2_1+H^2_2+H^2_3 + H_1H_2+ H_2H_3+ H_1H_3
\right)
\eeq
and
\beq
\label{kr}
R_{abcd}R^{abcd} = 4\left( 
\left( \dot{H}_1+H^2_1 \right)^2 + 
\left( \dot{H}_2+H^2_2 \right)^2 + 
\left( \dot{H}_3+H^2_3 \right)^2 +
H^2_1H^2_2 +
H^2_1H^2_3 +
H^2_2H^2_3 
\right).
\eeq
We have
\begin{eqnarray}
\frac{1}{a^2_1}G_{11} &=& \dot{H}_2 +\dot{H}_3 + H^2_2+H^2_3+H_2H_3 \\
\frac{1}{4a^2_1}Q_{11} &=& 2\dddot{H}_1+4\left( H_1
+H_2+H_3\right)\ddot{H}_1 \nonumber \\
& &+\left(3\dot{H}_1+2\dot{H}_2+2\dot{H}_3-2H_1^2
+8H_1H_3+8H_1H_2+4H_2H_3\right)\dot{H}_1 \nonumber \\
& &  +\left(\dot{H}_2+2H_1^2+2H^2_2-4H_1H_2\right)\dot{H}_2
 +\left(\dot{H}_3+2H_1^2+2H^2_3-4H_1H_3\right)\dot{H}_3   \\ 
& & -H_1^4+H_2^4+H_3^4+H_1^2\left( H^2_2+4H_2H_3+H^2_3\right) + 
H^2_2H^2_3 \nonumber \\
& &-2H_1\left( H_2^3 + H_3^3 + H^2_2H_3 + H_2H^2_3\right)\nonumber \\
\frac{1}{4a^2_1}S_{11} &=& 2\left( \dddot{H}_1+\dddot{H}_2+\dddot{H}_3
\right) 
+ 4\left( H_1+H_2+H_3\right) \ddot{H}_1 \nonumber\\
& & + 2\left( H_1+3H_2+2H_3\right) \ddot{H}_2
 + 2\left( H_1+2H_2+3H_3\right) \ddot{H}_3 \nonumber \\
& & +\left(3\dot{H}_1
+ 4\left(\dot{H}_2+\dot{H}_3 \right)
-2H_1^2+2\left(H_2+H_3\right)^2
+2H_1\left(H_2+H_3 \right) 
\right)\dot{H}_1 \nonumber \\
& & +\left(5\dot{H}_2+6H^2_2+4H_3^2+2H_1\left(H_2+H_3\right)
+8H_2H_3\right)\dot{H}_2  \nonumber \\
& & +\left(5\dot{H}_3+4H^2_2+6H_3^2+
2H_1\left( H_2+H_3\right)+8H_2H_3\right)\dot{H}_3  
+6 \dot{H}_2\dot{H}_3 \nonumber \\
& & -H_1^4 + H_2^4 +H_3^4 -2\left(H_2+H_3\right)H^3_1
+ 2H^3_2H_3 + 2H_2H_3^3 + 3H_2^2H_3^2 \nonumber \\
& & - \left( H_2+H_3\right)^2H_1^2.
\end{eqnarray}
The other nonvanishing components ${}_{22}$ and ${}_{33}$ are obtained 
by index cyclic
permutations from the above ones. One can check that in the isotropic
case $H_1=H_2=H_3=H$, we have $Q_{11}=Q_{22}=Q_{33}=2\dddot{H}+
12H\ddot{H}+ 9 \left(\dot{H}+2H^2 \right)\dot{H}
$ and $S_{11}=S_{22}=S_{33}=6\dddot{H}+
36H\ddot{H}+ 27 \left(\dot{H}+2H^2 \right)\dot{H}$, in such a way that all the contributions from the higher order
term cancel out in Einstein equations and we stay with the
original equations of the model \cite{PRD}.

We notice that due the term (\ref{kr}), we do not have anymore the
first integral $q/r=(H_1-H_2)/(H_2-H_3) =$ constant. This first integral
was identified in \cite{PRD2}, and it is a consequence of the
internal symmetry of Hilbert-Einstein action for 
Bianchi I metric (\ref{metric}) described in \cite{Chimento}.
Contrary to the the Kretschman scalar $R_{abcd}R^{abcd}$, the
scalar curvature $R$ is preserved under linear combinations
of $H_i$ that preserve the quantities $P=H_1+H_2+H_3$ and
$S=H_1^2+H_2^2+H_3^2$. The intersection of constant $P$ and $S$
corresponds to a circumference in the Euclidean 
space of $H_i$, and its $SO(2)$ 
symmetry is the responsible for the first integral\cite{Saa}.

With the contribution from the tensors $Q_{ab}$ and
$S_{ab}$, the singular equations (\ref{eq}) and (\ref{er}) are replaced
by the regular ones
\begin{eqnarray}
\label{eqddd}
\alpha\left(4\dddot{q} + \dots \right)+ F(\phi)\dot{q} = \left( 
F(\phi)p + F'(\phi)\dot{\phi}
\right)q,  \\
\label{erddd}
\alpha\left(4\dddot{r} + \dots \right)+ F(\phi)\dot{r} = \left( 
F(\phi)p + F'(\phi)\dot{\phi}
\right)r.
\end{eqnarray}
Both equations (\ref{eqddd}) and (\ref{erddd}) are free from
singularities on the surface $F(\phi)=0$. With the hypothesis
of small $\alpha$, the new terms are relevant only when 
$F(\phi)$ vanishes. They assure a regular behavior of the solutions
$q(t)$ and $r(t)$, eliminating the singular behavior present in
the original equations (\ref{eqddd}) and (\ref{erddd}).

\section{Conclusion}
 
 We shown that the inclusion of the conformal counterterm
 $C_{abcd}C^{abcd}$ in the action (\ref{act}) can eliminate the
 anisotropic singularity corresponding to $F(\phi)=0$. Such
 singularity is present in the recently proposed
 quintessential model \cite{PRD}, and it was its strongest
 objection. The conformal properties of the Weyl tensor $C_{abcd}$
 ensure that all isotropic solutions of (\ref{act}) are preserved
 when the counterterm is included. Hence, all the interesting
 dynamical behavior described in \cite{PRD} is still valid,
 including a cosmological history suitable, in principle, 
 for describing quintessence. 

A relevant issue is the naturalness of the necessary
adjust in the constants $\alpha_1, \alpha_2$ and $\alpha_3$
in order to have the conformal counterterm. As this constants
may arise from quantum corrections, the only reasonable
hypothesis about them is that they must be small if compared
with $1/G$. This point is now under investigation\cite{work},
and preliminary results show that, under the only hypothesis
of small $\alpha_1, \alpha_2$ and $\alpha_3$, both singularities
on $F(\phi)=0$ and $F_1(\phi)=0$ can be eliminated preserving
almost all of the isotropic behavior.

\acknowledgements

The authors acknowledge the financial support from the EEC 
(project HPHA-CT-2000-00015) and FAPESP (Brazil).


\begin{references}
\bibitem{PRD} E. Gunzig, A. Saa, L. Brenig, V. Faraoni, T.M. Rocha Filho, and 
A. Figueiredo, Phys. Rev. {\bf D63}, 067301 (2001); 
Int. J. Theor. Phys. {\bf 40}, 2295 (2001). 
%%CITATION = GR-QC 0012085;%%
%%CITATION = GR-QC 0012105;%%

\bibitem{PRD2} L.R. Abramo, L. Brenig, E. Gunzig. and A. Saa,
Phys. Rev. {\bf D67} 027301 (2003); Int. J. Theor. Phys. {\bf 42} 
1145 (2003).
%%CITATION = GR-QC 0210069;%%
%%CITATION = GR-QC 0210090;%%

\bibitem{G}G. Esposito-Farese and D. Polarski, Phys. Rev. {\bf D63},
063504 (2001).
%%CITATION = GR-QC 0009034;%%


\bibitem{Starobinski} A. A. Starobinski, P. Astron. Zh. {\bf 7}, 67 (1981)
[Sov. Astron. Lett. {\bf 7}, 36 (1981)]. 

\bibitem{Futamase} T. Futamase and K. Maeda, Phys. Rev. {\bf D39}, 
399 (1989); T. Futamase, T. Rothman, and R. Matzner,
 Phys. Rev. {\bf D39}, 405 (1981).
%%CITATION = PHRVA,D39,399;%%
%%CITATION = PHRVA,D39,405;%%



\bibitem{s2} S. Deser, Phys. Lett. {\bf 134B}, 419 (1984);
Y. Hosotani, Phys. Rev. {\bf D32}, 1949 (1985);
O. Bertolami, Phys. Lett. {\bf 186B}, 161 (1987).
%%CITATION = PHLTA,B134,419;%%
%%CITATION = PHRVA,D32,1949;%%
%%CITATION = PHLTA,B186,161;%%


\bibitem{Birrel} N.D. Birrel and P.C.W. Davies,
{\em Quantum fields in curved spaces}, Cambridge University
Press, 1982.

\bibitem{Buchbinder}I.L. Buchbinder, S.D. Odintsov, and
I.L. Shapiro, {\em 
Effective action in quantum gravity}. IOP Publishing ,1992.

\bibitem{Zeldovich}Ya.B. Zeldovich, Sov. Phys. JETP Lett. {\bf 6},
883 (1967).

\bibitem{Parker} L. Parker, Phys. Rev. Lett. {\bf 21}, 562(1968);
Phys. Rev. {\bf 183}, 1057 (1969); {\bf D3}, 346 (1971).
%%CITATION = PRLTA,21,562;%%
%%CITATION = PHRVA,183,1057;%%
%%CITATION = PHRVA,D3,346;%%

\bibitem{HartleHu} 
B.L. Hu and L. Parker, Phys. Rev. {\bf D17},
933 (1978); {\bf D17}, 3292 (1978);
M.V. Fischetti, J.B. Hartle, B.L. Hu,
{\bf D20}, 1757 (1979);  J.B. Hartle, B.L. Hu,
{\bf D20}, 1772 (1979);  {\bf D21}, 2756 (1980); 
%%CITATION = PHRVA,D17,933;%%
%%CITATION = PHRVA,D20,1757;%%
%%CITATION = PHRVA,D20,1772;%%
%%CITATION = PHRVA,D21,2756;%%


\bibitem{VCDM} L. Parker and A. Raval, 
Phys. Rev. Lett. {\bf 86},749 (2001);
Phys. Rev. {\bf D60}, 063512 (1999);
{\bf D60}, 123502 (1999); {\bf D62}, 083503 (2000);
{\bf D67}, 029901 (2003);
{\bf D67}, 029902 (2003);
{\bf D67}, 029903 (2003);
L. Parker and D. A.T. Vanzella, Phys.\ Rev.\ D {\bf 69}, 104009 (2004).  
%%CITATION = PRLTA,86,749;%%
%%CITATION = GR-QC 0003103;%%
%%CITATION = GR-QC 9908013;%%
%%CITATION = GR-QC 9905031;%%
%%CITATION = GR-QC 0312108;%%



\bibitem{Shapiro}I.L. Shapiro and J. Sola, JHEP {\bf 0202}, 006 (2002);
I. L. Shapiro, J. Sola, C. Espana-Bonet, P. Ruiz-Lapuente,
Phys. Lett. {\bf 574B}, 149 (2003). 
%%CITATION = HEP-TH 0012227;%%
%%CITATION = ASTRO-PH 0303306;%%


\bibitem{Chimento} L.P. Chimento,
Phys. Rev. {\bf D68}, 023504 (2003) 
%%CITATION = GR-QC 0304033;%%

\bibitem{Saa} A. Saa, work in progress.
\bibitem{work}E. Gunzig and A. Saa, work in progress.


\end{references}
\end{document}